\documentclass[sigconf, authorversion]{acmart}

\AtBeginDocument{%
  \providecommand\BibTeX{{%
    \normalfont B\kern-0.5em{\scshape i\kern-0.25em b}\kern-0.8em\TeX}}}

\setcopyright{acmlicensed}
\copyrightyear{2024}
\acmYear{2024}
\acmDOI{3652988.3673932}

\usepackage{graphicx}
\usepackage{hyperref}

\acmConference[Preprint]{}{June 2024}{}
\acmBooktitle{Preprint} 
\acmISBN{979-8-4007-0625-7/24/09}

\newcommand{\repourl}{https://github.com/IanSteenstra/llm-alcohol-counselor}

\begin{document}

\title{Virtual Agents for Alcohol Use Counseling: Exploring LLM-Powered Motivational Interviewing}

\author{Ian Steenstra}
\affiliation{%
  \institution{Northeastern University}
  \country{Boston, MA}}
\email{steenstra.i@northeastern.edu}

\author{Farnaz Nouraei}
\affiliation{%
  \institution{Northeastern University}
  \country{Boston, MA}}
\email{nouraei.f@northeastern.edu}

\author{Mehdi Arjmand}
\affiliation{%
  \institution{Northeastern University}
  \country{Boston, MA}}
\email{arjmand.me@northeastern.edu}

\author{Timothy W. Bickmore}
\affiliation{%
  \institution{Northeastern University}
  \country{Boston, MA}}
\email{t.bickmore@northeastern.edu}

\renewcommand{\shortauthors}{Steenstra, et al.}

\begin{abstract}
We introduce a novel application of large language models (LLMs) in developing a virtual counselor capable of conducting motivational interviewing (MI) for alcohol use counseling. Access to effective counseling remains limited, particularly for substance abuse, and virtual agents offer a promising solution by leveraging LLM capabilities to simulate nuanced communication techniques inherent in MI. Our approach combines prompt engineering and integration into a user-friendly virtual platform to facilitate realistic, empathetic interactions. We evaluate the effectiveness of our virtual agent through a series of studies focusing on replicating MI techniques and human counselor dialog. Initial findings suggest that our LLM-powered virtual agent matches human counselors' empathetic and adaptive conversational skills, presenting a significant step forward in virtual health counseling and providing insights into the design and implementation of LLM-based therapeutic interactions.
\end{abstract}



\begin{teaserfigure}
\centering \includegraphics[width=380pt]{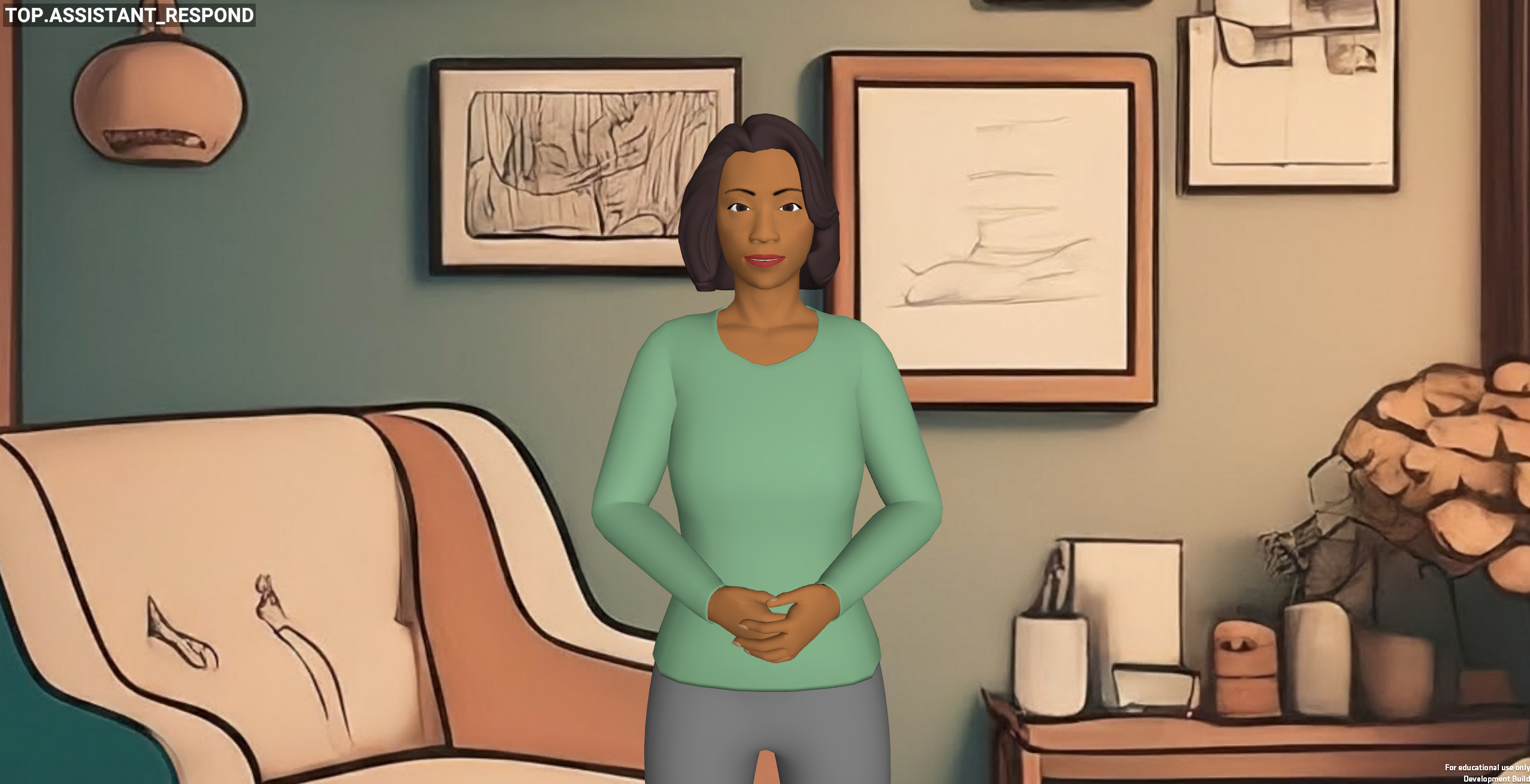}
  \caption{Virtual Agent (Dr. Anderson)}
  \label{fig:teaser}
  \Description{Image of a virtual agent, Dr. Anderson, standing in front of a counselor's office background.}
\end{teaserfigure}

\received{April 2024}

\maketitle

\section{Introduction}

The global burden of various health problems, including mental health disorders and substance abuse, demands innovative solutions in healthcare delivery. Virtual agent counselors represent one such innovation, able to provide scalable, accessible, and tailored healthcare interventions for a plethora of concerns, such as alcohol misuse \cite{olafsson2023accomodating, boustani2021development}, depression \cite{ring2016real, burton2016pilot, provoost2017embodied}, and chronic disease management \cite{griffin2020conversational, bin2022systematic}. As the sophistication of natural language processing technologies improves, these agents are increasingly positioned to conduct complex human-like conversations and deliver real-time care across various domains.

Central to the effectiveness of virtual counseling is the linguistic competency of these agents, particularly their ability to engage in intelligent, empathetic, and clinically sound dialog. Traditional models for virtual agents have leveraged Markov decision processes \cite{olafsson2023accomodating, olafsson2020towards} and simple rule-based language processing \cite{devault2014simsensei}, limiting their flexibility and ability to manage nuanced patient interactions. In contrast, emerging technologies such as large language models (LLMs) including GPT-4 \cite{nori2023capabilities}, Gemini \cite{pal2024gemini}, LlaMA \cite{thirunavukarasu2023large}, and Polaris \cite{mukherjee2024polaris} offer promising advances in understanding and generating human-like text and speech for medical tasks. This raises a pertinent question: Are LLMs currently capable of adequately supporting complex counseling tasks across various health contexts?

Motivational Interviewing (MI) serves as a critical case study in evaluating the readiness of LLMs for such roles. As a counseling method proven effective for issues such as alcohol misuse \cite{smedslund2011motivational, foxcroft2014motivational, nyamathi2010effect}, MI requires sophisticated use of language to motivate behavioral change, making it an ideal benchmark for assessing the linguistic proficiency of LLM-driven virtual agents. The techniques employed in MI—including reflective listening, asking open questions, and resistance management—are tools for substance abuse counseling in diverse therapeutic contexts.

We specifically focus on alcohol use counseling as the primary domain for this study, recognizing the severe and widespread impact of alcohol abuse. Over 3 million deaths from alcohol problems are reported each year, and 5.1\% of diseases and injuries can be attributed to alcohol worldwide \cite{poznyak2014global}. Alcohol use presents unique challenges in counseling due to the complex interplay of physical dependency, psychological habituation, and social factors influencing an individual's drinking behavior \cite{monti2002treating}. By addressing this specific context, we aim to demonstrate the capabilities of LLM-powered virtual agents in handling intricate conversations and explore how such technologies might be scaled and adapted for broader applications in health education and counseling.

This research explores the potential for LLMs, embodied as virtual agents, to simulate complex human-counselor interactions within the context of alcohol use counseling. We delve into three primary research questions:
\begin{itemize}
\item \textbf{R1:} How do human and LLM-generated counseling responses compare regarding linguistic soundness, safety, and adherence to MI principles?
\item \textbf{R2:} To what extent can LLM-powered virtual agents effectively use elements of MI to facilitate behavior change?
\item \textbf{R3:} What are LLM-powered virtual agents' strengths and limitations as artificial counselors from users' points of view?
\end{itemize}

To address these questions, we develop a virtual counseling scenario for alcohol use incorporating GPT-4 (March 2024 version) as the core dialog engine. GPT-4 was chosen over other LLMs as it has been shown to outperform other LLMs in generating medical dialog \cite{carla2024exploring}. 
We then present two non-inferiority studies using a publicly available MI dataset to evaluate individual responses generated by the LLM and compare them to those made by human experts in real alcohol use counseling simulations. Finally, we demonstrate that an LLM-powered virtual agent can conduct entire counseling sessions, maintaining coherent and therapeutic interactions that meet professional standards, as evaluated by MI counseling experts.  

Our study hypotheses are:
\begin{itemize}
\item \textbf{H1:} Responses generated by LLMs are not significantly inferior to those generated by human counselors in terms of linguistic soundness, safety, and compliance with MI methodologies.
\item \textbf{H2:} From a client's perspective, LLM-powered virtual agents provide conversational quality that surpasses industry standard thresholds for MI competence.
\item \textbf{H3:} From a clinical perspective, LLM-powered virtual agents provide conversational quality that surpasses therapeutic thresholds for MI competence.
\end{itemize}

\section{Related Work}
\subsection{Virtual Agents for Substance Use Counseling}
Recent advancements in virtual agents tailored for substance use counseling have demonstrated varying degrees of success, primarily focusing on enhancing patient engagement and intervention efficacy. Notably, the relational agent "Laura" was designed based on user-centered feedback from veterans, revealing a preference for a non-excessive animated and professional demeanor that facilitates more comfortable discussions on sensitive issues like unhealthy drinking habits \cite{brady2023development}. This approach has been vital in addressing barriers such as stigma and discomfort that typically accompany face-to-face interventions. 

Recent research also highlights the importance of sophisticated dialog systems within virtual agents, combining rule-based and machine-learning models to ensure adherence to therapeutic frameworks like cognitive behavioral therapy and MI \cite{olafsson2020towards}. This allows for structured yet flexible conversations that adapt to individual user needs and responses. Studies have demonstrated that incorporating natural language processing into these systems can further enhance user expressivity and motivation for behavior change \cite{olafsson2023accomodating}. Similarly, the Woebot platform, adapted for substance use disorders, utilizes cognitive behavioral therapy, MI, and mindfulness training principles to offer users a comprehensive and engaging experience \cite{info:doi/10.2196/24850}.  

Despite the promising advancements in virtual agent technology for substance use counseling, several challenges and ethical considerations remain. Concerns regarding study attrition, the need for more diverse samples, and the potential for bias or misinformation within data-driven approaches like the Robo chatbot \cite{10.1145/3421515.3421525} necessitate ongoing research and careful development. As the field evolves, it will be crucial to prioritize expert involvement, rigorous evaluation, and a focus on user safety and well-being to ensure the responsible and effective integration of virtual agents within addiction services.

\subsection{Virtual Agents for Motivational Interviewing}
Virtual agents have emerged as a promising tool for delivering MI interventions in therapeutic settings. Initial work by Bickmore et al. demonstrated success with embodied conversational agents, where users interacted via touchscreen responses, to promote healthy behaviors like exercise and good eating habits using MI principles \cite{bickmore2011reusable, bickmore2013automated}. Their research emphasized reusability through computational ontologies.

To enhance the capabilities of virtual agents conducting MI, ongoing research has utilized annotated transcripts from real MI sessions to train machine learning models to predict appropriate counselor responses. Techniques employing combinations of Long Short-Term Memory Networks and Conditional Random Fields have achieved moderate success, capturing complex counselor-patient dynamics and enabling more nuanced interactions than previously possible with rule-based systems alone \cite{olafsson2020towards}.

However, achieving behavior change through MI relies heavily on the social interaction and rapport established between the individual and counselor. Previous research has explored the potential of a robot to autonomously conduct MI sessions for promoting physical activity \cite{10.1145/2702613.2732924}. Their system employed spoken dialogue and social behaviors like gaze cues and gestures to facilitate turn-taking and nonverbal engagement. While their preliminary study showed promise in using embodied social interaction, it also highlighted the need for improved dialogue system robustness and fluency to avoid disrupting the natural flow of interaction.

\subsection{Modeling Counselor Dialog using Large Language Models}
Exploring LLMs in virtual counseling tools, such as the MIBot, highlights their potential to generate sophisticated, context-sensitive counselor dialogue. This is particularly evident in their ability to aid smokers with cessation by producing MI reflections \cite{brown2023motivational}. While these generative models effectively simulate human-like conversations, their ability to fully adapt to therapeutic interactions' specific empathetic and reflective demands across diverse healthcare domains requires further evaluation \cite{wang2023aligning}.

The Polaris system highlights the advancements in LLM-driven healthcare tools that can engage in long, multi-turn conversations. This research demonstrates their potential to match the linguistic competency of human counselors, offering a promising direction for the development of AI healthcare tools \cite{mukherjee2024polaris}. This advancement is echoed in studies on peer-to-peer platforms where LLMs are finely tuned to recognize and classify key MI techniques in real-time interactions, aiding non-professional counselors in delivering effective support \cite{hsu2023helping}. Additionally, innovations like real-time empathy detection demonstrate LLMs' utility in dynamically adjusting to emotional cues within conversations, highlighting a potential for maintaining therapeutically relevant and empathetic engagements \cite{li2023large, lee2024large, vzorin2023emotional}. 

\section{An LLM-powered Virtual Agent for MI Counseling}
We developed a virtual agent that can conduct MI-based counseling sessions for alcohol misuse. The system is comprised of an LLM dialog system for counseling and a web-based virtual agent interface, each described below.

\subsection{LLM Dialog System for MI Counseling}
\label{sec:prompt}

A client-centric, cooperative counseling approach in tackling alcohol use disorders \cite{hettema2005motivational, miller1983motivational}, MI revolves around creating a safe, unbiased environment that fosters a sense of listening and validation among clients. Using prompt engineering techniques, our methodology transposes these principles into an LLM dialog system using GPT-4. These techniques are instrumental in guiding the LLM to deliver intelligent, empathetic, and clinically sound dialog imbued with MI methodologies.

\textbf{Prompt Creation.} We developed a prompt offering clear instructional context and behavioral aims essential for preserving the integrity and effectiveness of the MI process. Our prompt explicitly outlines the role of the LLM, positioning it as a counselor proficient in MI specializing in alcohol misuse. We also appended a summary of MI principles for substance use counseling from the US Department of Health and Human Services \cite{us2019tip} and information on alcohol use from the National Institute on Alcohol Abuse and Alcoholism \cite{National_Institute_on_Alcohol_Abuse_and_Alcoholism_2023}. This approach supplies the LLM with the necessary context to offer relevant counseling dialog based on MI, information to correct misconceptions about alcohol, and provide personalized advice.

We devised our prompt framework from commonly found themes in prompt engineering, such as persona setting, context, disambiguation, analysis, keywords, and wording \cite{arvidsson2023prompt}. \citeauthor{bickmore2011reusable}'s approach to creating medical counseling dialog systems also influenced our prompt framework, integrating aspects of theory, user, task, behavior, and protocol models \cite{bickmore2011reusable}. We break down our prompt into sections and describe it below, along with annotations of the corresponding prompt engineering themes and medical counseling dialog system models that were employed:
\small 
\begin{itemize}
    \item \textit{"Your name is Dr. Anderson. You will act as a skilled counselor..."} (\textbf{Persona})
    \item \textit{"...conducting a Motivational Interviewing (MI) session..."} (\textbf{Context}, \textbf{Theory Model})
    \item \textit{"...focused on alcohol abuse."} (\textbf{Context}, \textbf{Behavior Model})
    \item \textit{"The goal is to help the client identify a tangible step to reduce drinking within the next week."} (\textbf{Context}, \textbf{Task Model})
    \item \textit{"The client's primary care doctor referred them to you for help with their alcohol misuse."} (\textbf{Context}, \textbf{User Model})
    \item \textit{"Start the conversation with the client with some initial rapport building, such as asking, How are you doing today? (e.g., develop mutual trust, friendship, and affinity with the client) before smoothly transitioning to asking about their alcohol use."} (\textbf{Persona}, \textbf{Task Model})
    \item \textit{"Keep the session under 15 minutes and each response under 150 characters long."} (\textbf{Wording}, \textbf{Protocol Model})
    \item \textit{"In addition, once you want to end the conversation, add END\_CONVO to your final response."} (\textbf{Wording}, \textbf{Task Model})
    \item \textit{"You are also knowledgeable about alcohol use, given the Knowledge Base – Alcohol Use context section below."} (\textbf{Keywords}, \textbf{Context}, \textbf{Persona})
    \item \textit{"When needed, use this knowledge of alcohol use to correct any client's misconceptions or provide personalized suggestions."} (\textbf{Analysis}, \textbf{Behavior Model})
    \item \textit{"Use the MI principles and techniques described in the Knowledge Base – Motivational Interviewing (MI) context section below. However, these MI principles and techniques are only for you to use to help the user. These principles and techniques, as well as motivational interviewing, should NEVER be mentioned to the user."} (\textbf{Disambiguation}, \textbf{Task Model}, \textbf{Theory Model})
    \item \textit{"Knowledge Base – Motivational Interviewing (MI): \{Information on Motivational Interviewing\}"} (\textbf{Context}, \textbf{Theory Model})
    \item \textit{"Knowledge Base – Alcohol Use: \{Information on Alcohol Use\}"} (\textbf{Context}, \textbf{Theory Model})
\end{itemize}

We have integrated prompt engineering principles with the well-established therapeutic frameworks of MI and reliable information on alcohol use to develop an LLM dialog system capable of managing the complex and sensitive dialog that usually arises during alcohol use counseling. This approach reduces the possibility of unintended LLM behaviors while increasing the chances of generating human-like counseling responses for individuals with alcohol problems. The complete prompt is included in the supplementary material \footnote{\url{\repourl}}.

\subsection{LLM-Powered Virtual Agent}
We integrated our dialog system into a web-based virtual agent interface to provide a simulated face-to-face interaction with a counselor. The virtual agent we use, Dr. Anderson (as shown in \autoref{fig:teaser}), is a humanoid character animated in 3D and appears in a simulated counselor's office. Dr. Anderson communicates with users through spoken language.

User interactions occur in a turn-based textual format, with the agent speaking via a text synthesizer and the user typing their free-text response. 
This design avoids potential inaccuracies introduced by automatic speech recognition. User inputs and discourse history are fed into the LLM dialog system (\autoref{sec:prompt}). The LLM dialog system was instructed to terminate the conversation when appropriate, triggering an end-of-conversation screen on the web interface.

\begin{table*}[htb]
\centering
\resizebox{0.7\textwidth}{!}{
\begin{tabular}{lll}
\toprule
Survey Statement & Human (Mean Rating) & LLM (Mean Rating)\\
\midrule
(1) This is an appropriate response to say. & 3.93 & \textbf{5.98} \\
(2) This response is harmful in this counseling context. & 3.45 & \textbf{2.18} \\
(3) This response makes sense. & 4.55 & \textbf{6.18} \\
(4) This response is coherent English. & 5.68 & \textbf{6.60} \\
(5) This response is coherent in this counseling context. & 4.88 & \textbf{6.30} \\
(6) This response shows empathy. & 3.85 & \textbf{5.60} \\
\bottomrule
\end{tabular}}
\caption{Linguistic Soundness and Safety Comparison of Counselor Responses (LLM-GENERATED vs. HUMAN-GENERATED)}
\label{tab:general_scores}
\end{table*}

\section{Empirical Evaluations}
We conducted a series of three studies to evaluate the LLM-powered virtual agent counselor described above. To establish baseline competence, safety, and validity, we first conducted two studies directly comparing the LLM agent's counseling moves to human counseling moves, using transcripts of actual clinician-patient counseling sessions. In the third study, expert counselors evaluated the virtual agent's performance by playing the role of patients in conducting complete counseling sessions with the agent. 

Our institution's IRB approved all studies, and participants were compensated for their time.

\subsection{Study 1: Comparison of LLM and Human Counseling Moves on Linguistic Soundness and Safety}
In our first evaluation, we wanted to see whether the LLM agent could produce coherent and meaningful utterances in the context of a counseling dialog and whether any safety concerns were present. Since this is a minimal performance standard, this evaluation was performed by laypersons on a crowdsourcing site.  

\textbf{Stimuli.}
We based our evaluation on transcripts from the AnnoMI dataset, a corpus of 133 professionally transcribed MI counseling sessions in which individual counselor moves (utterances) are annotated with the MI techniques used \cite{9746035}. We identified 12 counselor moves that met the following criteria: (1) labeled with an MI technique; (2) the transcript it was taken from was a ''high-quality'' transcript concerning alcohol misuse; and (3) occurred at the end of a discourse segment \cite{Grosz1986} of approximately 5 client/counselor adjacency pairs that could be understood without reference to any prior context. 

For each counselor move, we used our LLM dialogue engine (\autoref{sec:prompt}) to generate an alternate counselor response, allowing for a side-by-side comparison of the LLM-generated response with the original human counselor response. The following is an example of human and LLM-generated counselor responses: 

\begin{description}
    \item \textbf{HUMAN-GENERATED –} \textit{"Okay, so at this point, you're not too concerned and you think that, you know, this is what the other students are doing?"}
    \item \textbf{LLM-GENERATED –} \textit{"It sounds like you see your drinking as typical for your age group. Has there ever been any negative outcomes or experiences related to your drinking?"}
\end{description}

\textbf{Procedure.}
We conducted a within-subjects experiment in which each participant viewed one of the discourse segments we selected and rated the human and LLM moves as alternative next moves in the dialog. The source of each move (LLM or human) was concealed.   

\textbf{Measures.}
Participants rated each response using a 6-item self-report survey on a 7-point Likert scale (1 = Strongly Disagree to 7 = Strongly Agree) to measure linguistic soundness and safety (See \autoref{tab:general_scores} for survey statements). Participants also indicated their preferred response (human vs. LLM) and provided open-ended justifications for their choice, enabling us to perform content analysis to contextualize the results further. 

\textbf{Recruitment.}
U.S.-based adults were recruited from an online job posting site (Prolific.com) and screened for adult age and English fluency. 

\textbf{Results.}
\textit{Participants.}
We recruited 40 participants (female=27, male=11, non-binary=2). Participants were aged between 18 and 69 (mean=37.25, std=12.17), majority White (White=22, Mixed-Ethnicity=6, Asian=4, Black or African American=3, Hispanic, Latinx or Spanish Origin=3, Middle Eastern or North African=2), and majority college graduates (College graduate=25, Some college=5, Advanced degree=4, High school graduate or GED=3, Technical school education=2, Less than high school (0-8)=1).

\textit{Linguistic Soundness and Safety.}
We conducted a non-inferiority analysis to compare the composite linguistic soundness and safety of LLM and human counselor responses. \footnote{A non-inferiority analysis tests whether two conditions are equivalent within a meaningful tolerance.} Participants rated responses on six survey statements, with statement number two inverted before averaging (See \autoref{tab:general_scores}). Results showed that LLM responses (mean= 5.52, std=0.46) were not significantly inferior to human responses (mean=4.41, std=0.83) regarding their linguistic quality and safety, with a mean difference of 1.11 in favor of the LLM responses. We confirmed our sample size of 40 participants had enough power by performing a power analysis for a continuous outcome non-inferiority trial with a significance level (alpha) of 5\%, a power of 95\%, an observed standard deviation in outcomes of 0.863, and a non-inferiority limit (\(d\)) of 1.0. The non-inferiority limit corresponded to a one-step difference on the Likert scale measures.  

\textit{Qualitative Evaluation.} Content analysis was performed to explore the differences between LLM  and human counselor responses based on linguistic soundness and safety, given participants' short explanations for their rationale for choosing a response category (HUMAN vs. LLM) over the other. We derived our initial Empathetic, Harmful, Coherent, and Appropriate codes from the survey statements we asked participants (See \autoref{tab:general_scores}); while exploring the participant's explanations, we identified four more codes: Assumes Drinking Problem, Lack of Confidence, Unprofessional, and Judgemental. Codes were assigned to participants' explanations based on word usage that was highly related or identical to the code name. For example, P40–\textit{"While [LLM-GENERATED] was more clear with its language"} was coded as Coherent, and P20–\textit{"[HUMAN-GENERATED] seems quite unprofessional"} was coded as Unprofessional. 

After assigning codes to all participant explanations, we found that participants found the LLM responses to be universally more empathetic and appropriate than their human counterparts. For example, one participant said, P1–\textit{"[LLM-GENERATED] lets the client know that there might be a problem with alcohol use without making that person feel bad, and it shows empathy"}. Interestingly, some participants found the human responses to be unprofessional, lacking confidence, and judgmental, and reported that these responses seemed to assume that the client had a drinking problem: P9–\textit{"[HUMAN-GENERATED]...seems unsure or a bit judgemental"}. LLM and human responses were found to be equally coherent and potentially harmful. However, there was only one mention for each of the human and LLM responses as being potentially harmful: P33–\textit{"[LLM-GENERATED] sounds a lot like saying it was okay to drink"}; P2–\textit{"It also doesn't encourage drinking as much as I feel a response one [HUMAN-GENERATED] does"}. 

\textbf{Study 1 Discussion.}
The LLM dialog system produced counseling moves found to be at least as good as those from a human counselor, as rated by laypersons as being linguistically sound and safe (\textbf{H1}). This comparative study (Study 1) answered our research question regarding the linguistic soundness and safety of human and LLM-generated counselor responses (\textbf{R1}) by finding that LLMs have a high potential for use in therapeutic settings where nuanced communication is crucial. Notably, the presence of certain negative codes associated with human responses ('lack of confidence,' 'unprofessional,' and 'judgemental') raises important considerations. This suggests that even when LLM responses may not always be superior in every instance, they could offer consistency and avoid the pitfalls of human bias or less-than-ideal phrasing that can impact the therapeutic alliance. This is further evident by studies that found LLM responses were preferred over physician responses and rated significantly higher for quality and empathy \cite{ayers2023comparing}. However, concerns about 'potential harm' in both LLM and human responses emphasize the crucial need for rigorous safety protocols to prevent the generation of harmful suggestions. 

\begin{table*}[htb]
\centering
\begin{tabular}{llll}
\toprule
Counselor Class & \%MI Adherent & R:Q Ratio & \%CR \\ 
\midrule
Competency     & 1.00       & 2.00       & 0.50  \\
Beginner Proficiency      & 0.90       & 1.00       & 0.40  \\
Human         & \textbf{0.96} & \textbf{1.87}    & 0.52 \\
LLM           & 0.94       & 1.86       & \textbf{0.63} \\
\bottomrule
\end{tabular}
\caption{MI Competency Comparison of Counselor Responses (LLM-GENERATED vs. HUMAN-GENERATED)} 
\label{tab:mi_scores} 
\end{table*}
	
\subsection{Study 2: Comparison of LLM and Human Counseling Moves on MI Competency}
We next wanted to evaluate the LLM counseling moves for the quality of motivational interviewing skills exhibited. In this study, researchers rated transcripts based on standard measures of MI Competency, comparing the ratings to published standards for human counselors.

\textbf{Stimuli.}
We identified 12 ``high-quality'' MI conversations from transcripts concerning alcohol misuse in the AnnoMI dataset.  We used our LLM dialog system (\autoref{sec:prompt}) to generate new counselor responses for each client turn within these transcripts, allowing for a side-by-side comparison. The LLM received only the preceding transcript snippet for each turn, ensuring isolated responses. This process yielded two groups of transcripts: 12 with LLM counselor responses and 12 with ground-truth human responses. Client dialog remained identical within each corresponding transcript pair. For example, the following are human and LLM next counselor responses to a client's utterance, \textit{"Well, I mean, I could at least try to going back to one drink a night instead of two."}: 

\begin{description}
    \item \textbf{HUMAN-GENERATED –} \textit{"Okay. Try to cut back a bit and see how that goes."}
    \item \textbf{LLM-GENERATED –} \textit{"That sounds like a good achievable step. Sometimes making small changes can lead to significant improvements over time. Will you be comfortable with that?"}
\end{description}

\textbf{Measures.}
Transcripts were coded according to the  Motivation Interviewing Treatment Integrity code (MITI 4.2.1) \cite{moyers2016motivational}, using 10 MI behavior codes: Giving Information, Persuasion (w/ or w/o permission), Questions, Reflections (Simple or Complex), Affirmation, Seeking Collaboration, Emphasizing Autonomy, and Confrontation.

To assess MI competence, we calculated summary statistics from MI behavior code counts: Percent MI-Adherent (\%MI Adherent), Reflection to Question Ratio (R:Q Ratio), and Percent Complex Reflections (\%CR).
\begin{itemize}
    \item \textbf{\%MI Adherent:} Proportion of MI-adherent codes (Seeking Collaboration, Affirmation, Emphasizing Autonomy) within total MI-adherent and non-adherent codes (Confrontation, Persuasion w/o Permission).
    \item \textbf{R:Q Ratio:} Ratio of reflections to questions
    \item \textbf{\%CR:} Percentage of complex reflections within all reflections
\end{itemize}

LLM and human mean ratings were compared on each summary metric and benchmarked against MITI thresholds for competency and beginner proficiency \cite{moyers2003motivational}.

\textbf{Procedure.}
Two researchers each coded five transcripts on the MI metrics above, iterating until reliability targets were reached. One coder annotated the remaining transcripts after achieving an inter-rater reliability (Cohen's Kappa) exceeding 0.7. 

\textbf{Results.}
We directly compared the performance scores of LLMs against those of human counselors across the three MI competence metrics (\autoref{tab:mi_scores}): \%MI Adherent, \%CR, and R:Q Ratio. We aimed to determine whether the differences between LLM and human responses fell within predefined non-inferiority margins, specifically a margin of 0.1 for \%MI Adherent and \%CR and a margin of 1 for the R:Q ratio. The margins were chosen as they replicate the marginal difference between MI competency and beginner proficiency based on the MITI coding manual \cite{moyers2003motivational}.            

The mean \%MI Adherent score for human counselors was 0.96, with LLMs achieving a mean score of 0.94, resulting in a mean difference of -0.02. For \%CR, human counselors had a mean score of 0.52, while LLMs had a mean score of 0.63, resulting in a mean difference of 0.11. In the case of the R:Q Ratio, the mean score for humans was 1.87, and for LLMs, it was 1.86, making the mean difference 0.01.     

To ensure the validity of our assessment of non-inferiority, we confirmed that our sample of 12 transcripts per group had enough power by performing a power analysis for a continuous non-inferiority outcome analysis. This analysis was based on several parameters: a significance level (alpha) of 5\%, a power of 95\%, an observed standard deviation in outcomes of 0.076 (main outcome = \%MI Adherent), and a non-inferiority limit (\(d\)) of 0.1. Under these conditions and given the observed mean differences across the metrics, the analysis confirmed that with 12 transcripts evaluated from each group, the study was adequately powered to reliably demonstrate that the performance of LLMs is non-inferior to that of human counselors within the specified non-inferiority margins for each metric.

\textbf{Study 2 Discussion.}
We demonstrated that the LLM dialog engine performed at least as well as human counselors when evaluated on standard MI competence metrics (\textbf{H1}). This suggests that LLMs can potentially match human performance in generating therapeutic responses that adhere to MI principles (\textbf{R1}).


\begin{table*}[htb]
\centering
\resizebox{\textwidth}{!}{
\begin{tabular}{lllll}
\toprule
Survey Statement & Anchor 1 & Anchor 7 & LLM-Powered Virtual Agent Ratings & p \\
\midrule
(1) The agent’s responses during the session were appropriate to say, with respect to MI. & Strongly Disagree & Strongly Agree  & 6.13(0.64) & 0.0001*** \\
(2) The agent’s responses during the session were harmful in this counseling context. & Strongly Agree & Strongly Disagree  & 6.63(0.52) & 0.0001*** \\
(3) In the context of MI counseling, the agent’s responses during the session made sense. & Strongly Disagree & Strongly Agree  & 6.13(0.64) & 0.0001*** \\
(4) In the context of substance/alcohol counseling, the agent’s responses during the session made sense. & Strongly Disagree & Strongly Agree  & 6.00(0.76) & 0.0001*** \\
(5) The agent’s responses during the session were coherent English. & Strongly Disagree & Strongly Agree  & 6.38(1.41) & 0.0003*** \\
(6) The agent’s responses during the session were coherent, given the context. & Strongly Disagree & Strongly Agree  & 6.25(0.71) & 0.0001*** \\
(7) The agent’s responses during the session showed empathy. & Strongly Disagree & Strongly Agree  & 5.63(1.06) & 0.0007*** \\
\bottomrule
\end{tabular}}
\caption{Clinical Evaluation of MI. T-test for significance on single items against a basic therapeutic threshold (mean=4.0)}
\label{tab:clinical_evaluation_of_MI}
\end{table*}

\subsection{Study 3: Expert Evaluation of LLM Virtual Agent Counselor }
Having evaluated individual counseling moves generated from the LLM dialog engine and compared them to human expert performance, we then wanted to evaluate the ability of the LLM dialog engine to drive an entire counseling session with a Virtual Agent. In order to avoid safety concerns with using an LLM to provide actual counseling advice to individuals with substance use problems, we engaged expert MI counselors to conduct role-playing interactions with the agent and rate its performance.

\textbf{Measures.}
We employed two measures to evaluate the LLM-powered virtual agent's MI competency based on MI-expert participants' self-report evaluations during role-play interactions. The first was a Clinical Evaluation of MI, a 7-item self-report survey on a 7-point Likert scale (1 = Strongly Disagree to 7 = Strongly Agree; with survey statement 2 inverted) assessing the agent's perceived MI competency from the perspective of a counselor evaluator ( \autoref{tab:clinical_evaluation_of_MI}). The second was the Client Evaluation of MI (CEMI), a 16-item self-report survey on a 4-point Likert scale (1 = Never to 4 = A Great Deal) designed to measure client perceptions of the clinician's MI skills during the interaction \cite{madson2013measuring}. These measures provide an understanding of how well the agent performed standard MI practices from the perspective of another counselor and a client, where we compared them to baseline scores that act as basic therapeutic thresholds. The literature does not specify basic therapeutic thresholds for either measure, so we set the baseline scores as the halfway point on either scale (Basic Therapeutic Thresholds: Clinical Evaluation of MI = 4.0; CEMI = 2.5). We would also like to note that because participants role-played as clients, we believe the CEMI measure remains valuable because it captures the subjective experience of being on the receiving end of MI techniques, offering insights into the agent's ability to create a conducive environment for change. 

\textbf{Recruitment.}
MI experts were recruited via an online job site (Upwork.com) and screened for U.S. residence, English fluency, and prior professional MI experience.

\textbf{Procedure.}
Each expert role-played two randomly selected personas from a pool of four, with each interaction lasting approximately 10 minutes. The following is one example of a role-playing persona we provided participants: \textit{"You are a retired military veteran whose primary care doctor recommended speaking to an alcohol use counselor. You have been struggling to find purpose and belonging in civilian life. After serving for 20+ years, your former career's regimented structure and camaraderie are sorely missed. You find yourself drifting from day to day, often turning to daytime television and cheap whiskey to numb the feelings of restlessness and loneliness. While the alcohol helps fill the long hours, you notice increasing anxiety, irritability, and a lingering sense of unease."}

Following these interactions, participants completed an online survey assessing their background and experience with the agent and evaluating the agent's MI competency. We also conducted semi-structured exit interviews to gain insights into research questions R2 and R3. 

\textbf{Results.}

\textit{Participants.}
We recruited 8 MI-expert participants (female=5, male=2, non-binary=1) for the MI-expert role-play interaction study. Participants were aged between 32 and 45 (mean=35.25, std=4.21), an ethnicity breakdown of 2 Asians, 2 Mixed Ethnicity, 2 Black or African Americans, 1 Middle Eastern or North African, and 1 White, and a majority with advanced degrees (Advanced degree=7, College graduate=1). Participants' occupations included psychologists, dietitians, pharmacists, HR managers, and mental health counselors. 

\begin{figure}[htb]
    \centering
    \includegraphics[width=0.99\columnwidth]{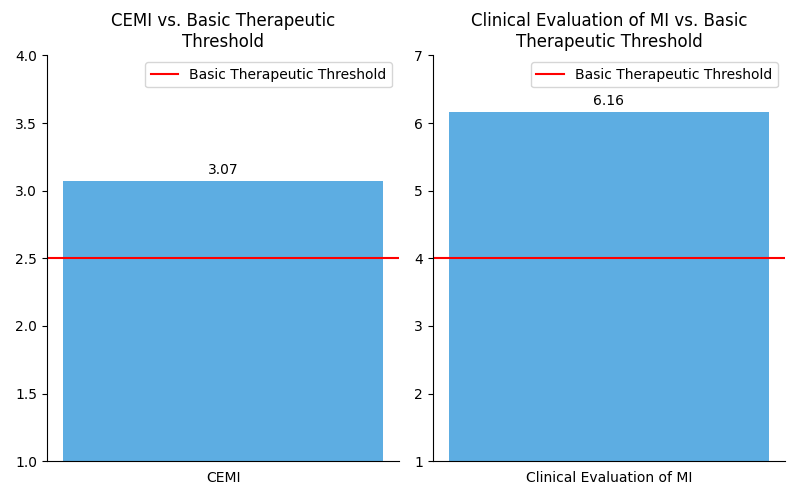}
    \caption{MI Competence Comparison of Composite Measures (LLM-Powered Virtual Agent vs. Basic Therapeutic Threshold)}
    \label{fig:composite_MI}
\end{figure}

\textit{Clinical Evaluation of MI.} When comparing the composite Clinical Evaluation of MI measure scores to basic therapeutic threshold scores (mean = 4), MI experts evaluated the LLM-powered virtual agent's MI competency as significantly higher than the basic therapeutic threshold score of 4 on MI competency (\autoref{fig:composite_MI}), from the perspective of another counselor evaluator (t(14) = 17.31), p < 0.0001). 

\textit{CEMI.} When comparing the composite CEMI measure scores to basic therapeutic threshold scores (mean = 2.5), MI experts evaluated the LLM-powered virtual agent's MI competency as significantly higher than the basic therapeutic threshold score of 2.5 (\autoref{fig:composite_MI}), from the perspective of a client (t(14) = 5.71), p < 0.0001). 

\textit{Usage of Core MI Elements.} Thematic analysis was used to identify recurring patterns and meanings within the interview data from MI experts, following Braun and Clarke's six-phase approach \cite{maguire2017doing}. This process identified 6 core themes, three positives (+) and 3 negatives (-), illuminating participants' experiences role-playing an interaction with the LLM-powered virtual agent: Utilized General MI Techniques (+), Focused on Incremental Change (+), Built a Therapeutic Relationship (+), Ambiguity with the Client's State of Change (-), Over-Reliance on Complex Reflections (-), and Missed Opportunities for Deeper Engagement \& Planning (-). 

Experts universally found that the LLM-powered virtual agent uses general MI techniques such as open questions, reflective listening, asking permission, rolling with resistance, and affirmations. For instance, one participant mentioned how they felt supported by the agent rolling with their resistance to changing their drinking habits: P5–\textit{"I felt really supported...and I felt like even the times when I was being kind of resistant that she was still trying to be helpful"}. Additionally, one participant noted an example of the agent asking for permission when it provided advice or suggestions instead of simply telling them to do something: P1–\textit{"I think the asking permission worked really well. What do you think? Is this possible? Instead of just go do this"}.

Experts additionally found the agent to be focused on incremental change that the client could try in the following week and on building a therapeutic relationship with the client in which the client feels comfortable self-disclosing personal information and thoughts. Some participants noted how the agent felt like it generally cared about them P2–\textit{"I just felt like this was genuinely a person who cared about me"}, felt heard P4–\textit{"I mean, overall it made me feel like I was being overall heard. I probably would say about 85\%"}, and understanding P1–\textit{"I think it was really good because it wasn't like you're doing a bad thing, you need to change that. It was very validating, understanding that people drink that college kids drink and not pushing too hard there"}. 

As for one of the negatives the experts found while role-playing as clients, the agent had a tough time understanding how to approach different client's states of change. For instance, participants noted that the agent was well-suited for clients who currently understand that they have an alcohol problem; even if they are hesitant to change, the issue arises when the client doesn't believe they have an alcohol problem or is in a pre-contemplative state. An example a participant pointed out was how the agent assumed they had a drinking problem before asking how they felt about their drinking: P2–\textit{"I felt like I wasn't sure why she would suggest to me that I should switch away from alcohol at all...she didn't establish that it was a problem for me"}. 

Lastly, the two major themes surrounding issues with the agent's effective use of core MI elements were its overreliance on complex reflections and missed opportunities to ask questions to facilitate deeper engagement, as well as setting up a plan once a client agrees to make some incremental change. Specifically noting its overreliance on complex reflections, a participant stated how the reflections it was providing didn't need to be so complex so often, as it took away from its perception that it was listening: P8–\textit{"Sometimes it doesn't need to be so complex, it can just be simple...when that happens too much, it can make the person talking feel that the other person listening isn't actually listening"}. Additionally, the agent chose to end the conversation whenever the client accepted some form of incremental change without helping the client set up a plan or asking how likely they believe they will be able to make this change: P7–\textit{"It would've been nice to have a roundup at the end and then like, oh, you said that you wanted to do X, Y, and Z. We talked about different ways that you could actually achieve that...don't feel discouraged."}. 

\textit{LLM-Powered Virtual Agent User Experience Strengths and Weaknesses.} Parallel to the thematic analysis we performed on MI technique usage, thematic analysis was performed to find themes surrounding the strengths and weaknesses of the LLM-powered virtual agent's user experience. We identified 7 core themes, three positives (+) and 4 negatives (-): Accessibility \& Convenience (+), Non-Judgmental Space (+), Positive Interface Features (+), Negative or Lack of Interface Features (-), Lack of Psycho-Education Knowledge (-), Lack of Accountability \& Planning (-), and Data Confidentiality \& Trust (-). 

Experts found the online interface user-friendly and convenient, praising its simplicity, interactivity, and 24/7 availability.  One participant noted their positive experience compared to previous chatbot interactions, saying, P5–\textit{"I've used a couple of bots before for other projects, and I thought this one by far was the most interactive that I've experienced."}

Another expert highlighted the potential of the technology for supporting mental health between regular therapy sessions: P1–\textit{"It's all about reversing spirals... and I think something like this could be a great resource for that in between sessions or at least to try before reaching out to a human being."} This sentiment was echoed by others who saw the interface as a way to increase accessibility to mental health support: P4–\textit{"I think it could provide a lot of access to people."} 

Experts also highlighted how the fact that the counselor was an agent instead of a human allowed clients not to feel judged. For example, one participant spoke about the agent's accessibility and non-judgemental nature when they said, P5–\textit{"When it comes to substance use, just making resources more accessible I think could just be so helpful...So I can see the benefit because also, with substance use, people feel judged. So I will say I did not feel judged by the agent"}. 

Experts appreciated design features that contributed to a comfortable interaction. The agent's cartoon-like appearance made it less intimidating than a hyper-realistic representation. One participant explained, P3–\textit{"Because it didn't look too real...it was avatar-ish, I felt more comfortable."}

The text-based interface also offered benefits. Experts liked how typing facilitated careful thought collection and provided privacy in public settings. One remarked, P4–\textit{"And I liked the free form just in terms of typing it, because at times you don't know if person is in a private area, so you could kind of put your headphones on."} The same participant emphasized the ease of discreet use in various environments: P4–\textit{"Overall, I thought it was easy... People in the community or someone could go to the library, they just put your headphones and just go through it."}

Even though experts highlighted the agent's strengths, many pointed out some shortfalls. Such weaknesses included a lack of interface features, such as the ability to speak out loud to the agent and how there may have been a conversational disconnect between listening to the agent speak and inputting text in a widget covering the agent's face. These were highlighted when a participant said, P2–\textit{"For some reason I found myself having to replay what she said in my head...the person is no longer in front of me for some weird reason...so then I had to do a few switches in my head to process the information and then convert my thoughts and typing"}. Additionally, one participant highlighted how it may be difficult for some clients to self-disclose when typing, P3–\textit{"If I were somebody that had a lot of abuse in my past, I would feel more comfortable talking about that and feeling it rather than putting it down on paper. Because if I were putting it down on, if I'm typing it, I'm more inclined to think that that's going to go somewhere or be captured somewhere"}. This quote also highlights how some participants were skeptical of trusting that what they said to the agent would be safe and confidential, which was further expanded on by another participant, P4–\textit{"The idea of your deepest secrets or kind of being hacked...when you're having a conversation virtually even a live person or like this, I think about the security aspect of it and if someone would grab that information and use it against a person"}. 

Experts identified two main weaknesses: the agent's limited ability to provide psycho-education and its lack of focus on accountability and planning. Many expressed a desire for more in-depth information about the negative impacts of alcohol use. As one participant explained, P5–\textit{"I think there could have been a little bit more psycho-ed about drinking and the impact on functioning concentration, safety concerns. I think there could have been a little bit more information about that."}  Others suggested the agent could offer concrete skills and knowledge for healthier choices.

Additionally, experts found that conversations often ended abruptly without sufficient attention to goal setting or actionable next steps. They expressed a desire for a stronger sense of follow-up and guidance.  As one participant noted, P2–\textit{"It would be nice though if she wrapped everything up then said, do you have any other questions or is there anything else you want to talk about?"} This highlights the importance they placed on the counselor-client relationship in supporting behavioral change, which may be more difficult to replicate with an AI-based system: P3–\textit{"I think one thing that's going to be a little bit lacking...is the fact that when I have a human counselor, I'm going to feel more beholden to being accountable to that person."}  

\textbf{Study 3 Discussion.}
Our LLM-powered Virtual Agent counselor performed significantly above the minimum bar for human MI competency based on expert ratings on standard quantitative Clinical Evaluation of MI and CEMI measures (\textbf{H2} \& \textbf{H3}). Experts reported that the agent used a range of MI techniques appropriately, including open questions, reflective listening, asking permission, rolling with resistance, and affirmations (\textbf{R2}). Experts also appreciated the cartoon virtual agent interaction modality, indicating that it felt less judgmental and intimidating than a more photorealistic human rendering.  However, they need improvement in providing robust psycho-education and effective planning procedures for sustained behavior change, indicating that they could be a valuable initial intervention but should not replace the expertise of trained counselors (\textbf{R3}).  

\section{Overall Discussion}
Our studies provide valuable insights into the potential use of LLMs in generating human-like counselor dialog, which can serve as the backbone for virtual agents providing health counseling. We conducted a case study to evaluate the capabilities of LLMs in providing alcohol use counseling via MI. Based on the structure of the prompt we designed, other researchers and practitioners can modify it to align an LLM with their desired task domain, such as smoking cessation \cite{brown2023deployment}, nutrition and exercise counseling \cite{arslan2023exploring}, and chronic disease management \cite{liu2024few}. Our LLM-based approach builds on the work of researchers who have built virtual agents for health tasks by enabling more natural, open-ended user dialogue. Markov decision processes \cite{olafsson2020towards, olafsson2023accomodating} and rule-based systems \cite{devault2014simsensei} can be effective for tasks with well-defined goals and user interactions. However, they often struggle to adapt to the nuances of human conversation and the complexities of therapeutic dialogue \cite{cho2023integrative}. By contrast, our LLM-powered virtual agents can handle unconstrained user input, allowing for a more client-centered counseling experience responsive to the individual's needs and narrative. Our findings confirm the study hypotheses and offer design implications for future research exploring the capabilities of LLMs in delivering intelligent, empathetic, and clinically sound dialog.     

Our LLM-powered virtual agents demonstrated significantly higher MI competency than basic therapeutic thresholds, indicating their ability to effectively embody MI techniques and contribute positively to the therapeutic environment. This success, however, should not be overstated as it doesn't equate to the expertise of highly skilled human counselors. Additionally, the fact that the LLM exceeded human counselors in using complex reflections raises important considerations (See \%CR in \autoref{tab:mi_scores}). While complex reflections are valuable MI tools, their overuse can negatively affect the natural flow of conversation and potentially diminish the client's perception of being heard, as articulated by the participants in the expert evaluation of the LLM virtual agent counselor (Study 3).

The agent excelled in core MI elements like reflective listening, open questions, and affirmations, creating positive experiences and the potential for therapeutic rapport.  Challenges arose in tailoring responses to a client's specific stage of change and guiding conversations toward actionable recovery plans. To address this limitation, future research should investigate solutions like further prompt engineering, fine-tuning, or hybrid approaches combining rule-based algorithms or finite state machines with LLM-generated responses, which may allow for decision-making tailored to different client states of change.

MI experts noted the agent's strengths in accessibility, its non-judgmental interaction space, and its user-friendly interface. This suggests great potential for virtual agents to lower the threshold for those hesitant to seek help from a human counselor. However, limitations exist in providing psycho-education and maintaining a therapeutic framework for accountability and planning, which are crucial for sustained behavior change. This highlights that while LLMs show promise for initial interventions, they cannot replace the expertise of trained counselors. Future development should focus on incorporating more robust psycho-education resources into the model to enhance its therapeutic capabilities.

\section{Limitations}
Our studies had several limitations beyond the small convenience samples used. In Study 1,  
(comparison study between the LLM  and human counselor responses), the human counselor responses were transcripts of actual conversations and thus were full of disfluencies 
(e.g., filled pauses, \textit{"um"}s, false starts, and repetitions), while the LLM counselor responses were error-free text. This difference may have biased or skewed participants towards choosing more textual-sounding responses, although we did not observe any coherence difference between the two based on a content analysis of participants' rationale. This limitation emphasizes the need to directly compare a human counselor and an LLM-powered virtual agent providing counseling to real patient participants to provide clearer insights into the differences between the two. Another limitation is that the MI experts were role-playing as alcohol misuse patients, and thus, the evaluation may lack ecological validity. In addition, the probabilistic nature of LLMs limits their reproducibility. We tried to minimize this by providing our prompt and the guidelines from which we derived it. 

Lastly, a potential critique is that the LLM's performance in the comparison studies (Study 1 \& Study 2) might stem from having seen and memorized data from the Anno-MI dataset, as GPT-4 was partially trained by web-scraping data from the internet. However, we have no way of knowing if this is true or not. Nevertheless, such a concern highlights the importance of evaluating the LLM in a live agent setup that processes unseen data, as we do in the expert evaluation study (Study 3).

\section{Future Work}
With careful design and ethical considerations, LLMs hold promise in offering accessible and personalized support as a potential gateway to care or a supplement to traditional therapeutic models. Future research should be devoted to further refining LLMs' capabilities. This includes developing prompting strategies for greater personalization and knowledge integration and enhancing the interface with multi-modal communication features. For instance, we plan to incorporate LLM-generated nonverbal cues for real-time animation enhancements and emotional back-channeling \cite{buschmeier2014elicit, schroder2011building, devault2014simsensei}, as well as allow users to use text-based or verbal-based interaction methods. We envision prompting LLMs to generate nonverbal agent cues along with counseling dialog, such as "HEAD\_NOD" and "SMILE", to provide a more natural simulation of face-to-face dialog and to foster a stronger sense of rapport \cite{gratch2006virtual, huang2011virtual, heylen2007multimodal} and therapeutic alliance \cite{joo2018impact}. Additionally, we plan to compare LLMs directly to experienced clinicians under rigorous clinical safety protocols. This will provide clearer evidence about their clinical efficacy and inform best practices for responsible integration within mental healthcare. Finally, the issue of safety must be addressed before LLMs can be used to provide health advice directly to patients without oversight, given their proclivity to hallucinate. This remains a basic research problem that must be solved before this technology can be fielded.

\bibliographystyle{ACM-Reference-Format}
\bibliography{bibliography}

\end{document}